\documentclass[aps,prl,showpacs,twocolumn,amsmath,amssymb]{revtex4-1}
\usepackage{graphicx,bbm,color}

\newcommand{\be}{\begin{equation}}
\newcommand{\ee}{\end{equation}}

\newcommand{\bra}[1]{{\langle #1 \vert}}

\newcommand{\ket}[1]{{\vert #1 \rangle}}

\newcommand{\ave}[1]{{\langle #1\rangle}}
\newcommand{\ii}{ {\rm i} }
\newcommand{\dd}{ {\rm d} }
\newcommand{\ZZ}{\mathbb{Z}}

\newcommand{\CC}{\mathbb{C}}

\newcommand{\x}{{\rm x}}
\newcommand{\z}{{\rm z}}

\newcommand{\mm}[1]{{\mathbf{#1}}}

\def\tr{{{\rm tr}}}
\def\ad{{\,{\rm ad}\,}}
\def\one{\mathbbm{1}}

\def\DD{\hat{\cal D}}
\newcommand{\La}{{\mathtt L}}
\newcommand{\Ra}{{\mathtt R}}

\begin{document}

\title{Exactly solvable counting statistics in weakly coupled open interacting spin systems} 

\author{Berislav Bu\v ca and Toma\v z Prosen}

\affiliation{Department of Physics, FMF,  University of Ljubljana, Jadranska 19, 1000 Ljubljana, Slovenia}

\date{\today}

\begin{abstract}
We study the full counting statistics for interacting quantum many-body spin systems weakly coupled to the environment. 
In the leading order in the system-bath coupling we derive exact spin current statistics for a large class of parity symmetric spin-1/2 systems driven by a pair of Markovian baths with local coupling operators.
Interestingly, in this class of systems the leading order current 
statistics are universal and do not depend on details of the Hamiltonian.
Furthermore, in the specific case of symmetrically boundary driven anisotropic Heisenberg ($XXZ$) spin 1/2 chain we derive explicitly the third-order non-linear corrections to the current statistics.
\end{abstract}

 
\maketitle

{\em Introduction.--}  In the past couple of decades substantial progress has been made in understanding the physics of nonequlibrium systems \cite{der, evans1, lebowitz, sch1, sch2}. These systems, after long times, evolve to steady states. One of the key aspects, and one may even say defining properties, of out-of-equilibrium statistical physics is the existence of macroscopic currents of charge, particles, heat, etc., which flow through the system even after the system has reached the nonequilibrium steady state. Although several methods exist for studying the average current in the steady state, it is much more challenging to study fluctuation properties of the current which in general depend not only on the (asymptotic) steady state of the system, but also on the correlations at earlier times. 

Apart from being challenging, studying the probability distribution of the current in various systems has attracted a lot of attention in recent years since it offers much more insight into the nature of the system studied than merely the average of the current \cite{rev}. Analytical results for nonequilibrium {\em interacting quantum} systems are particularly rare and progress has been made, only very recently, for some integrable spin chains in the Markovian approximation for the master equation describing driving of the system \cite{p11a, p11b, kps} and, even then, only the averages of the currents can be calculated by the methods used. Analytical results for non-integrable systems out-of-equilibrium are unheard of, even more so for the current statistics. Existing exact results for current statistics include small systems, such as quantum dots \cite{groth}, non-interacting systems such as Fermi gas and the $XX$ spin chain \cite{muz}, and, as of very recently, classical stochastic processes \cite{TASEP}, a Luttinger liquid conductor \cite{cond} and critical systems using conformal field theory \cite{cft}.

These results hold only for very specific systems. On the other hand, in this Letter we find a perturbative (in leading order of system-bath coupling) \emph{universal law for spin current statistics}, which holds for all nonequilibrium spin-1/2 systems, provided that they fulfil three requirements: (i) The coupling between the system and the baths (environment) is weak (and Markovian), (ii) and local. (iii) The systems fulfill a very weak condition of a parity-type symmetry (defined later). 
 
 The strongest of these requirements, the Markovian and locality approximations (i,ii), are justifiable if the system is weakly coupled to the environment and the interaction between the system and the environment is short range (which is, indeed, the case for spin interactions). Although still very strict (mostly due to the difficulty of isolating the environment degrees of freedom from the system degrees of freedom), the first two requirements are quite physical and experimentally realizable. For instance, important progress has been made recently in controlling the Markovianity of the time evolution of many systems \cite{mark, qsimul, qsimul2}. 
 Requirement (iii) is shown to hold in a wide range of systems, which, remarkably, include even (frustrated) spin chains with parity symmetric but possibly long-range and inhomogenous interactions. Such systems are more realistic than the integrable models usually studied (for instance see \cite{frust}). In fact, applicable systems even include multi-dimensional spin systems, such as spin lattices, systems studied in the context of coherent transport in photosynthetic complexes \cite{manzano1}, where the action of the environment is usually taken to act locally, and even other interacting complex networks with spin-like (or qubit) degrees of freedom \cite{majd}. Our results should therefore be particularly important for experimental realizations of spin systems with cold atoms and trapped ions { \cite{qsimul, qsimul2, mark} and possibly for quantum information transfer { \cite{cap}  and quantum computing,  where weakness of decoherence is crucial. We also note that the third cumulant of the current has been measured recently for certain systems \cite{third}. Using our leading order (in system-bath coupling) results helps us to also calculate explicit third-order (non-linear) system-specific corrections for the integrable case of the boundary-driven $XXZ$ spin 1/2 chain.

A powerful method which we use, first developed within quantum optics \cite{scully}, is the method of full-counting statistics \cite{levitov, rev}, based on introducing a \emph{counting field} which counts the number of times the system has had a unit of charge or spin pushed in a certain direction. The counting field allows one to compute all the moments of the current, which is equivalent to computing the full current distribution, and gives deep insight into the physical nature of the system studied.

{\em Full counting statistics in weakly coupled open spin-1/2 quantum systems.--}
Let us consider a system of $n$ spins $1/2$ desribed by Pauli operators $\sigma^\pm_j = \frac{1}{2}(\sigma^{\rm x}_j\pm\ii\sigma^{\rm y}_j),\sigma^{\rm z}_j,j=1,\ldots,n$ acting on a 
tensor product space $(\CC^2)^{\otimes n}$. 
We aim at computing the { full spin current statistics} in the limit of weak system-bath coupling when Markovian approximation for the system's density matrix $\rho(t)$ is appropriate \cite{open}.
The dynamics of the latter is then dictated by the systems's Hamiltonian $H$ and a set of Lindblad {\em jump} operators $L_m$, $m=1,2\ldots$ via the Lindblad equation
\be
\frac{\dd \rho(t)}{\dd t} = -\ii [H,\rho(t)] + \varepsilon \left ( \DD^{\rm jump} \rho(t) + \DD^{\rm diss}\rho(t) \right ),
\label{eq:lindblad}
\ee
where the non-unitary part of the generator is split, respectively, into {\em the quantum jumps} and {\em the dissipation}
\be
\DD^{\rm jump}\rho :=\sum_\mu L_m \rho L^\dagger_m,\quad
\DD^{\rm diss} \rho := \frac{1}{2}\sum_m \{ L^\dagger_m L_m,\rho\},
\ee
and $\varepsilon$ is a parameter describing the strength of system-bath coupling, which we will assume to be small enough to allow for applicability of perturbation theory. 
We shall consider the current of magnetization $M=\sum_{j=1}^n \sigma^\z_j$, which is assumed to be conserved by the Hamiltonian $[H,M]=0$.
\begin{figure}
 \centering	
\vspace{-1mm}
\includegraphics[width=0.85\columnwidth]{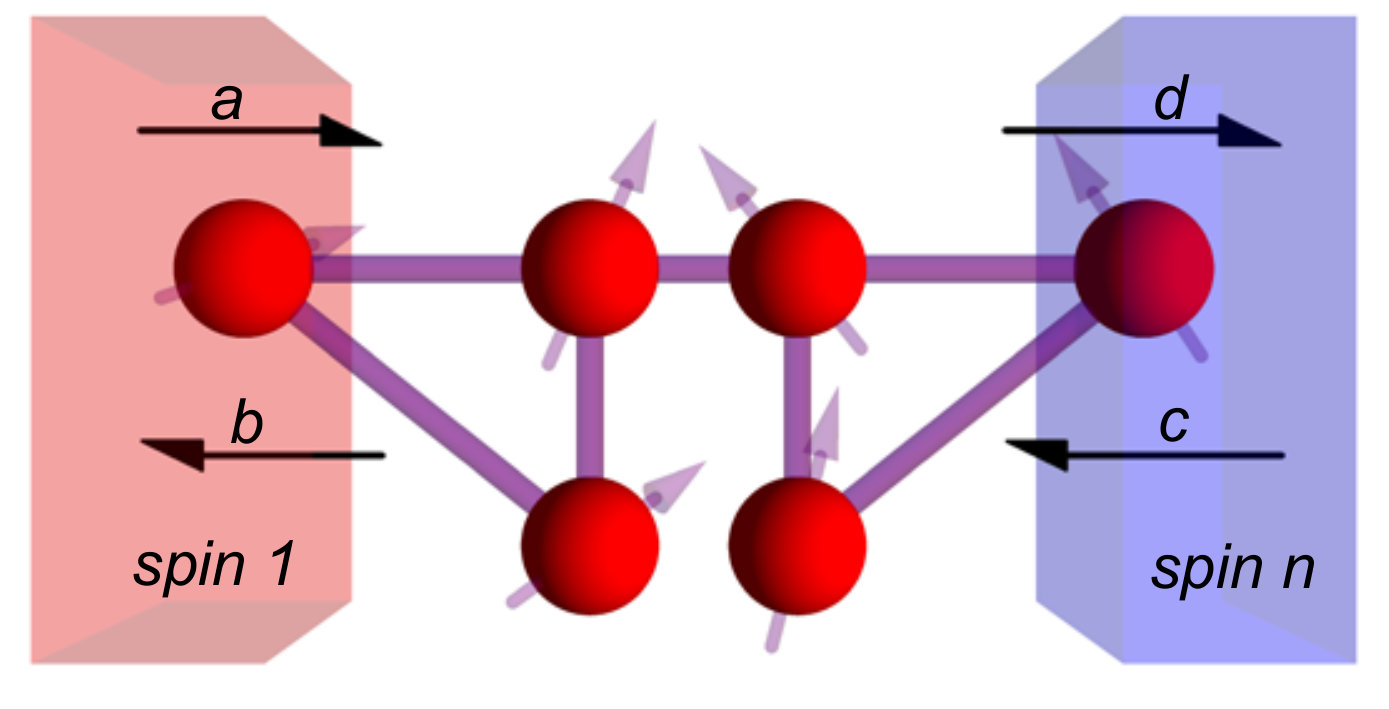}
\vspace{-1mm}
\caption{(Color online) The spin transport model: Exemplar parity-symmetric system is coupled locally to two spin baths (represented by boxes), at spin sites 1 and $n$. The baths act on the system via jump operators (\ref{gendriv}) with the corresponding rates ($a,b,c,d$). The arrows beneath the jump rates  indicate the direction in which the spin current is being driven.  }
\label{fig:fig1}
\end{figure}
We consider the most general form \cite{notegenform} of jump operators acting {\em locally} on a pair of sites only  
\begin{align}
L_{+,+} &= \sqrt{ a} \sigma^+_1, \quad L_{+,-} = \sqrt{ b} \sigma^-_1, \quad \nonumber \\
L_{-,+} &= \sqrt{ c} \sigma^+_n, \quad L_{-,-} = \sqrt{ d} \sigma^-_n,  \label{gendriv}
\end{align}
changing magnetization by $\pm 1$, $[M,L_{\mu,\nu}] = \nu L_{\mu,\nu}$, for $\mu,\nu\in\{\pm\}$,
where $a,b,c,d$ represent incoherent transition rates for the two coupled spins (see Fig.~\ref{fig:fig1}). By suitably adjusting $\varepsilon$, we shall fix $a+b+c+d=2$.
We argue that our jump operators represent a general model of transport between a {\em pair of baths}, attached to $j=1$ and $j=n$ spins,
provided the bath dynamics are fast compared to the dynamics generated by $H$ on initial excitations localized on sites $1$ or $n$, which is the condition ensuring ultra-locality of 
operators $L_m$ (\ref{gendriv}) in derivation of the master equation \cite{open}.
Let us denote the amount of quantity $M$ transported in time $t$ from the first bath to the second bath  by $N(t)$. Intuitively, we may see that the Lindblad jump operators acting on site 1, $L_{+, \nu}$, will either drive spin from the first bath into the system ($L_{+,+}$) or from the system into the bath ($L_{+,-}$), and similarly for $L_{-, \nu}$ acting on site $n$. Hence, $L_{\mu, \nu}$ with positive $\mu \nu = +$ will drive the current in the positive direction and those with negative $ \mu \nu = -$ in the negative direction.} In fact, in the open system's framework, $N(t)$ is exactly the sum of the times the Lindblad operators acting on site 1, $L_{+, \nu}$ inject spin from the first bath to the system minus the number of times they inject spin out of the system and back into the first bath in time $t$, and conversely for site $n$. In the steady state (long time) limit the currents calculated at site 1 and site $n$ are the same.

In this limit $I=\lim_{t\to \infty}\frac{1}{2 t} N(t)$ then equals the current of $M$, and the factor 2 in the denominator comes from counting the flow twice, on site 1 and on site $n$.
The statistics of currents can be asymptotically, for $t\to\infty$, fully characterized by the growth rates of cumulants of $N(t)$, which can be elegantly formulated \cite{rev, groth} in terms of introducing a {\em counting field} $\chi$ into the jump superoperator $\DD_\chi^{\rm jump}\rho :=\sum_{\mu,\nu} e^{\ii \mu \nu \chi} L_{\mu,\nu} \rho L^\dagger_{\mu,\nu}$

\be
\ave{I^m}_c := \lim_{t \to \infty}\frac{1}{2 t} \ave{[N(t)]^m}_c =  \frac {\partial^{m} \lambda(\chi) }{\partial (\ii \chi)^{m}} \Big |_{\chi \to 0}. \label{cs0}
\ee
Here $\lambda(\chi)$ is a leading eigenvalue (of maximal real part) of the modified Liouvillean
\be
\left [-\ii \ad H + \varepsilon \left ( \DD_\chi^{\rm jump} + \DD^{\rm diss} \right ) \right ]\rho(\chi) = \lambda(\chi)\rho(\chi) 
\label{eq:gen}
\ee
with $\rho(\chi):=\lim_{t \to \infty} \rho( \chi, t) $ the corresponding right eigenvector, introducing a superoperator $(\!\ad H)\rho \equiv [H,\rho]$ acting linearly on the space of operators.
Note that $\lambda(0) = 0$ and $\rho(0)$ is the non-equlibrium steady state density operator.

This method may be intuitively understood by observing a reduced density matrix $\rho_N(t)$, that is $\rho(t)$ projected to a subspace of $N$ spin-transfers between the two baths in time $t$. The trace of this, $P_N(t)=\tr \rho_N(t)$, is the probability of $N$ spin transfers in time $t$. By performing a Fourier transform (in $N)$ of this reduced density matrix, $\rho( \chi, t)=\sum_N \rho_N(t) e^{-\ii \chi N}$, it may then be shown (by observing the action of the generator of the time evolution) that the Lindblad master equation, Eq. \eqref {eq:lindblad}, has the jump superoperator  modified, $\DD_\chi^{\rm jump}\rho :=\sum_{\mu,\nu} e^{\ii \mu \nu \chi} L_{\mu, \nu} \rho L^\dagger_{\mu,\nu}$, so that it depends on the counting field $\chi$ and in which direction ($\mu \nu $) a specific Lindblad operator $L_{\mu,\nu}$ drives the flow. Furthermore,  if we normalize $\tr \rho (\chi, t=0)=1$, the largest eigenvalue of the Liouvillian, $\lambda (\chi)$, corresponds to the cumulant generating function for the current distribution in the long time limit \cite{levitov, rev}, since for large $t$, $\rho(\chi, t) \approx e^{\lambda(\chi) t} \rho(\chi,t= 0) $.

Let us turn to our problem and consider formal perturbation expansion of $\rho(\chi)$ and $\lambda(\chi) $ in the system-bath coupling strength $\varepsilon$, namely
 \be
\rho (\chi) = \sum_{p=0}^\infty (\ii \varepsilon)^p \rho^{(p)}, \quad \lambda (\chi)  = \sum_{p=1}^\infty \varepsilon^{2p-1} \lambda^{(2p-1)},
\ee
where all even orders of $\lambda$ vanish due to the fact that the current and all its cummulants should be odd functions of $\varepsilon$.
We may also normalize $\rho$ so that $\tr \rho^{(p)}=\delta_{p,0}$.
We will assume that the conditions of Evans theorem hold (or the Liouvillean can be symmetry reduced \cite{evans}) so the fixed point $\rho(\chi)$ is unique. 
The first two orders satisfy (defining $ \hat{\cal D}_\chi :=\DD_\chi^{\rm jump} + \DD^{\rm diss} $),
\begin{align} (\!\ad H) \rho^{(0)} &=0, \label{zo} \\ 
(\!\ad H) \rho^{(1)} + \hat{\cal D}_\chi \rho^{(0)}  &= \lambda^{(1)} \rho^{(0)}. \label{fo}
\end{align}
The zeroth order solution to \eqref{zo} can be formally written in terms of a full set of linearly-independent operators $Q_k$ that commute with $H$, $[H,Q_k]=0$,
$\rho^{(0)}=\sum_k \alpha_k Q_k/ (\tr \sum_k \alpha_k Q_k)$.
We may now make two general observations: (i) Taking the trace of \eqref{fo} we find  
\be
\lambda^{(1)}=\tr \hat{\cal D}_\chi\rho^{(0)}. \label{fos}
\ee
(ii)  Furthermore, to determine $\rho^{(0)}$ it is sufficient to require that $(\hat{\cal D}_\chi - \lambda^{(1)} )\rho^{(0)}$ is in the image space of the commutator, $\rm{Im} \ad H$, without actually solving Eq.~(\ref{fo}). We shall demonstrate below that this results in a very weak requirement for the Hamiltonian in case of two-gate coupling to the environment via local spin-flip jump operators.

{\em Leading order for spin-$1/2$ systems with two gates to environment.--} Guided by Ref.~\cite{p12} we make an ansatz for the zero-th order in terms of a magnetized infinite temperature equilibrium state,  
\be
\rho^{(0)} = 2^{-n}\prod_{j=1}^n(\one + \nu \sigma_j^{\rm z})
\label{eq:ansatz}
\ee with free parameter $\nu(\chi)$. Furthermore, we observe that since $\rho^{(0)}$ can be written in the form $\exp(\theta M)$ 
$[H,(\rho^{(0)})^{-1}]=0$, so multiplying by $(\rho^{(0)})^{-1}$ conserves the image space of $\ad H$, hence Eq.~(\ref{fo}) has a solution iff 
$(\rho^{(0)})^{-1}(\hat{\cal D}_\chi - \lambda^{(1)} )\rho^{(0)} \in {\rm Im}\ad H$. Using (\ref{fos},\ref{eq:ansatz}), the expression $(\rho^{(0)})^{-1}(\hat{\cal D}_\chi - \lambda^{(1)} )\rho^{(0)}$
is found to be a general combination of four terms $\one$, $\sigma^\z_1$, $\sigma^\z_n$, and $\sigma^\z_1 \sigma^\z_n$.
Requiring cancellation of the first and the last term results in two equations, determining uniquely $\lambda^{(1)}$ and $\nu$:
\begin{eqnarray}
\lambda^{(1)} &=& 2\sqrt{1+ (e^{-2 \ii \chi}-1)ad + (e^{2 \ii \chi}-1)bc}  - 2, \label{foa} \\
\nu&=& \frac{\lambda^{(1)}-(e^{-\ii \chi}-1)(a+d)+(e^{\ii\chi}-1)(b+c)}{(e^{-\ii\chi}-1)(a-d)+(e^{\ii\chi}-1)(b-c)}, \nonumber
\end{eqnarray}
while the remainder is a simple algebraic condition
\begin{equation}
-\sigma^\z_1 + \sigma^\z_n \in  {\rm Im}\ad H,
\label{eq:cond}
\end{equation}
which, if fulfilled, validates the initial simple ansatz (\ref{eq:ansatz}). 

We may now see that a parity-like symmetry is sufficient to ensure condition (\ref{eq:cond}), requiring the action of the dissipator on the equilibrium density operator of the system to be orthogonal -- in the Hilbert-Schmidt sense -- to all operators which commute with the Hamiltonian (i.e., it must be in the image of $\ad H$).} Namely, it is easy to show that (\ref{eq:cond}) holds generally for parity-symmetric Hamiltonians, i.e. if there exists an operator $P$, with $P^2=\one$, such that $P H = H P$, and satisftying at least one of the following properties
\begin{equation}
 \quad P \sigma^\z_1 = \sigma^\z_n P,\quad {\rm or}\quad  P \sigma^\z_{1,n} = -\sigma^\z_{1,n} P, \label{eq:Psym2}
\end{equation}
with additional {\em weak} requirement, namely that also all conserved operators, $Q_k$, e.g. as written in terms of eigenspace projectors, are parity symmetric $[P,Q_k]=0$ \cite{noteparity}.
Proof: Due to hermiticity of $\ad H$ w.r.t. Hilbert-Schmidt inner-product the condition (\ref{eq:cond}) is equivalent to: $\tr(-\sigma^\z_1+\sigma^\z_n)Q_k = 0, \forall Q_k$. Indeed:
$\tr(-\sigma^\z_1+\sigma^\z_n)Q_k = \tr P(-\sigma^\z_1+\sigma^\z_n)Q_k P =  -\tr (-\sigma^\z_1+\sigma^\z_n)Q_k$. 

We then apply Eq.~\eqref{foa} to calculate all the cumulants for this wide class of spin systems via Eq.~\eqref{cs0}. For instance, the expectation value of the spin current is $ \left <I_{(1)} \right >_c= \frac{\varepsilon}{2} (a d-b c)$.  { Closed form expressions for higher cumulants were obtained in the same way, but are lengthy and therefore we will not write them. However,} they significantly simplify if we consider a symmetric driving instead of a general one \eqref{gendriv}, 
\be
a=d=(1+\mu)/2, \qquad b=c=(1-\mu)/2,
\label{eq:symm}
\ee 
where the driving strength $\mu$ controls the nonequilibrium forcing due to unequal average spin polarizations of the two baths. Then we have $\nu=0$, $\rho^{(0)}=2^{-n} \one$, so $\lambda^{(1)}=-1+\cos\chi - \ii \mu \sin\chi $, and 
$\ave{I^{2k+1}_{(1)}}_c=\varepsilon \mu/2$ for odd cumulants and $\ave{I^{2k}_{(1)}}_c=\varepsilon/2$ for even cumulants. Extreme driving $\mu=1$ hence results in the Poisson distribution $\ave{I^m}_c={\rm const}$. 

Our results, stating that current statistics may not depend on details of $H$, hold generally only below a certain {\em perturbative border}, $\varepsilon < \varepsilon^*$, where clearly [see (\ref{eq:gen})] $\varepsilon^* \propto \|H\|$, so they can not be applied in the trivial case when one switches off the coherent interactions $H \to 0$.

{\em Explicit third order solution for the $XXZ$ chain.--} 
The preceding discussion for the leading order correction holds for all spin systems satisfying the parity-symmetry requirement. We may also find an explicit third order solution of the $XXZ$ spin chain with Hamiltonian

\be
H_{XXZ} = \sum_{j=1}^{n-1} (2\sigma^+_j \sigma^-_{j+1} + 2\sigma^-_j \sigma^+_{j+1} + \Delta \sigma^{\rm z}_j \sigma^{\rm z}_{j+1}),
\ee
and symmetric driving (\ref{eq:symm}), we are able to find the third order correction to the current cumulants by generalizing the solution for the steady state given in \cite{p11a}. $P$ is realized a permutation of spins $j \leftrightarrow n+1-j$, or as $\prod_{j=1}^n \sigma^\x_j$ \cite{p12b}. Solving Eq.~\eqref{fo} we find that the first order is now the same as without the counting field \cite{p11a}, $ 2^n \rho^{(1)}=c^{(1)} (Z-Z^\dagger)$ \cite{p11a}, but multiplied by a different constant, $c^{(1)}= (-\mu-\mu \cos\chi + \ii \sin\chi)/2 $. $Z$ is the pseudo-local conserved operator given in terms of a matrix product  in Ref.~\cite{p11a}.

Let us now turn to the second order equation reading 
\be
 (\!\ad H)\rho^{(2)} + \hat{\cal D}_\chi \rho^{(1)}= \lambda^{(1)} \rho^{(1)}.
 \label{eq:seco}
\ee
We { make an ansatz for} the second order solution, { similar in form} to the one for $\chi=0$ \cite{p11a}, namely, 
\be
2^n \rho^{(2)} = c^{(1)} c^{(2)}_1 (Z - Z^\dagger)^2 - c^{(1)}c^{(2)}_2 [Z,Z^\dagger]. 
\ee
The dissipator acts only on the two boundary sites. Thus we need to check the action of the dissipator on these sites only. We use $[H,[Z,Z^\dagger]]=(\hat{\cal D}_+ + \hat{\cal D}_-)(Z - Z^\dagger)$, and   $[H,(Z-Z^\dagger)^2] = -(\hat{\cal D}_+ - \hat{\cal D}_-)(Z - Z^\dagger)$, where,
$
 \hat{\cal D}_\pm\rho:= 2 \sigma^\pm_1 \rho \sigma^\mp_1 - \{\sigma^\mp_1 \sigma^\pm_1,\rho\}+
2 \sigma^\mp_n \rho \sigma^\pm_n - \{\sigma^\pm_n \sigma^\mp_n,\rho\},
$
as was shown in \cite{p11a}, to rewrite (\ref{eq:seco}) as, $
\Bigl[  c^{(2)}_1 (\hat{\cal D}_+\!-\!\hat{\cal D}_-)+c^{(2)}_2 (\hat{\cal D}_+\!+\!\hat{\cal D}_-)  + \hat{\cal D}_\chi -\lambda^{(1)} \Bigr] (Z\!-\!Z^\dagger)=0,$

which actually gives us six independent equations, only two of which turn out not to be redundant, and are solved by $c^{(2)}_1=\frac{1}{4}  (-\mu-\mu \cos\chi + \ii \sin\chi) $ and $c^{(2)}_2=\frac{1}{2} ( \cos\chi - \ii \mu \sin\chi)$.
The second order solution (of (\ref{eq:seco})) can be modified, however, by the addition of arbitrary conserved quantities, $Q_k$, where $[H,Q_k]=0$, namely, $\rho^{(2)'} =  \rho^{(2)} + \sum_k \alpha_k Q_k$. As we have checked by means of computer algebra, the existence of a solution to the third order equation, 
 \be
(\!\ad H) \rho^{(3)} + \hat{\cal D}_\chi \rho^{(2)'}=  \lambda^{(3)} \rho^{(0)}+\lambda^{(1)} \rho^{(2)'},
\ee
in fact requires nontrivial coefficients $\alpha_k$.
By taking the trace of this equation, we find that the third order correction to the current fluctuations reads 
 \be
 \lambda^{(3)} = \tr (\hat{\cal D}_\chi \rho^{(2)'}- \lambda^{(1)} \rho^{(2)'}). \label{tr3}
\ee
We can calculate $\lambda^{(3)}$ despite not knowing the full second order solution by observing several properties. Firstly, only terms of the form $O \in \{\one, (-\sigma^z_1+\sigma^z_n) \}$ in $\rho^{(2)'}$ can possibly contribute to \eqref{tr3}.  Secondly, because $ \tr (\hat{\cal D}_\chi \one)=\lambda^{(1)}\tr\one$, the contribution from $\one$ cancels out.
Finally, since $ (-\sigma^z_1+\sigma^z_n)$ is obviously in the image of the adjoint of the Hamiltonian (due to the existence of a solution to \eqref{fo}), the second order solution can not be modified by this term. 
We therefore have,
 \be
 \lambda^{(3)} = (-\mu+\mu \cos\chi + \ii \sin\chi)\tr[\bigl(-\sigma^z_1+\sigma^z_n) \rho^{(2)}\bigr] \label{tr3a}, 
\ee
which can be interpreted as, up to a constant, the drop in magnetization from one end of the chain to the other. 
\begin{figure}
 \centering	
\vspace{-1mm}
\includegraphics[width=0.85\columnwidth]{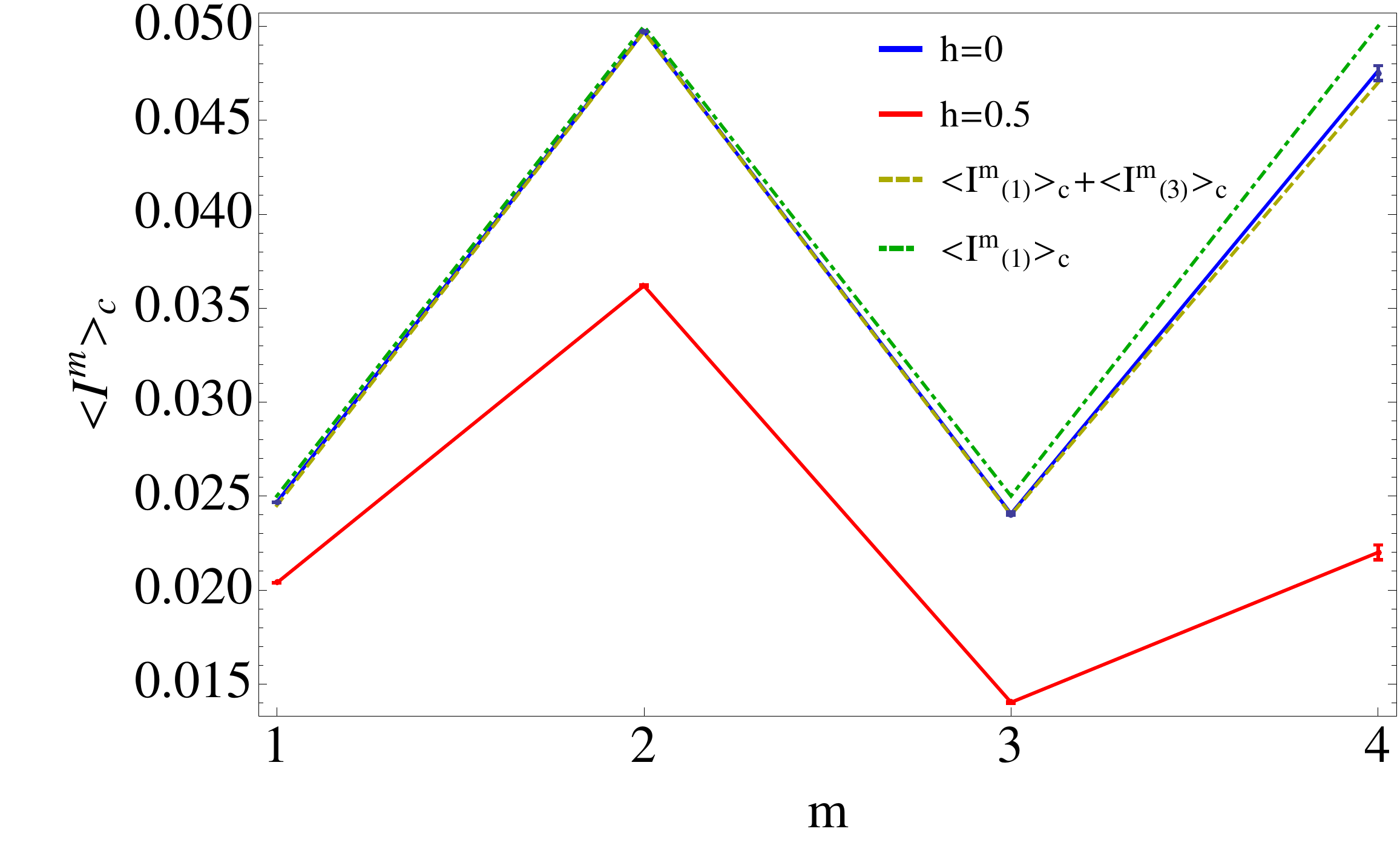}
\vspace{-1mm}
\caption{(Color online) The first four current cumulants obtained numerically for the $XXZ$ spin chain and staggered field $XXZ$ spin model, $H=H_{XXZ}+ \sum_{j=1}^{n}h  (\pm 1)^j  \sigma_j^{z}$, with field strength $h$, for $n=4$, $\Delta=0.5$, $\mu=0.5$, $\varepsilon=0.1$.
Dashed (chained) lines indicate analytical results up to second (fourth) order in $\varepsilon$. }
\label{fig:CF}
\end{figure}
The third order correction to the current fluctuations can now be computed similarly as before, yielding, separately for even/odd $m$, 
\begin{eqnarray}
\left <I^{2k}_{(3)} \right >_c &=-\varepsilon^3 f(n) \frac{(9^k-1) (3\mu^2+1)}{128 (2k)!},\quad \\
\left <I^{2k+1}_{(3)} \right >_c &= -\varepsilon^3 f(n)  \frac{ \mu ( 9^{k+1}-1 + 3 (9^k-1) \mu^2)}{256 (2k+1)!} \nonumber
\end{eqnarray}
where $f(n)={\bra{\La} \mm{T}^{n} \ket{\Ra}}-{\bra{\La} \mm{T}^{n-1} \ket{\Ra}}$ and $\mm{T}$ is exactly the transfer matrix from Ref. \cite{p11a}, acting on auxiliary Hilbert space with two ground-state vectors $\ket{\Ra}$ and $\ket{\La}$. This can be evaluated in a closed form for any anisotropy $\Delta$ of the form $\Delta = \cos(\pi l/ m)$, $l,m\in\ZZ$. For instance, for $\Delta=1$, we have, $f(n)=n-1$, and for $\Delta=1/2$, $f(n)=\frac{1}{45} (-1)^{-n} 8^{1-n} \left(5 (-8)^n-6 (-5)^n+10\right)$. We have checked our results numerically using the wave-function Monte Carlo method of quantum trajectories (see e.g. appendix of Ref.~\cite{qt1}) for the symmetrically driven $XXZ$ spin chain with $n=4$, $\Delta=0.5$, $\mu=0.5$, $\varepsilon=0.1$. The high order of our result allows for precise computations even at that not-so-small coupling, as shown in Fig. \ref{fig:CF}. We have also contrasted our example with a numerical simulation for the $XXZ$ spin chain in the staggered field which breaks the $P$-symmetry (\ref{eq:Psym2}) and for which the general results for $\ave{I^m}_c$ do not apply.
 
  {\em Discussion.--}
We have calculated the full current statistics, which does not depend on the details of the system's Hamiltonian, for a wide class of spin models with two-gate Markovian coupling to the environment up to the second-order in the system-bath coupling.  In the case of symmetric driving, Eq. \eqref{eq:symm}, our results are easy to appreciate: the spin flow, between the two baths (or leads), basically behaves as a biased random walk (with odd cumulants $ \ave{I^{2k+1}_{(1)}}_c=1/2 \epsilon \mu$ and even cumulants  $\ave{I^{2k}_{(1)}}_c=\varepsilon/2$), completely independent of the Hamiltonian.  Rather counter-intuitively, the Hamiltonian (which has to be strong compared to the dissipation) plays a marginally important role and is there only to ensure correlation between spins flowing in and out of the system via conservation of total magnetization, and the properties of the spin current depend only on the details of the weaker dissipation.  Apart from the universal leading-order result (in system-bath coupling), we have also found exactly the full current statistics for up to fourth-order for the integrable boundary driven Heisenberg $XXZ$ spin chain. We feel that the condition of {\em locality} could sometimes be weakened; that the degrees of freedom described by 
$\sigma^\alpha_1$ and $\sigma^\alpha_n$ need not be spatially localized, it is only important that they are simultaneously measurable.

%The details of this as well as extension to systems with larger degrees of freedom (e.g., higher spin systems or Hubbard models \cite{hub}) should be the topic of future research.

We acknowledge useful comments by B. \v Zunkovi\v c, E. Ilievski and M. \v Znidari\v c and support from the Grant P1-0044 of ARRS (Slovenia).


\begin{thebibliography}{10}
\bibitem{der} B. Derrida, J. Stat. Mech. ({\bf 2007}) P07023; {\em ibid.} ({\bf 2011}) P01030.

\bibitem{evans1} D. J. Evans, E. G. D. Cohen and G. P. Morriss, Phys. Rev. Lett. {\bf 71}, 2401 (1993); G. Gallavotti and E. D. G. Cohen, {\em ibid.} {\bf 74}, 2694 (1995); C. Jarzynski, {\em ibid.} {\bf 78}, 2690 (1997); K. Saito and A. Dhar, {\em ibid.} {\bf 99}, 180601 (2007).

\bibitem{lebowitz} J. L. Lebowitz and H. Spohn, J. Stat. Phys. {\bf 95}, 333 (1999).

\bibitem{sch1} B. Schmittmann and R. K. P. Zia, in `Phase Transitions and Critical Phenomena', edited by C. Domb and J. L. Lebowitz (Academic, San Diego, 1995), Vol. 17.

\bibitem{sch2} G. M. Sch\" utz, in `Phase Transitions and Critical Phenomena', edited by C. Domb and J. L. Lebowitz (Academic, San Diego, 2001), Vol. 19;  H. Spohn, Large Scale Dynamics of Interacting Particles (Springer-Verlag, New York, 1991).

\bibitem{rev}  M. Esposito, U. Harbola and S. Mukamel,  Rev. Mod. Phys. {\bf 81}, 1665�1702 (2009)

\bibitem{p11a} T. Prosen, Phys. Rev. Lett. {\bf 106}, 217206 (2011).

\bibitem{p11b} T. Prosen, Phys. Rev. Lett. {\bf 107}, 137201 (2011).

\bibitem{kps} D. Karevski, V. Popkov and G. M. Sch\" utz,  Phys. Rev. Lett. {\bf 110}, 047201 (2013); E. Ilievski and B. \v Zunkovi\v c, {\tt arXiv:1307.5546};  T.~Prosen, E.~Ilievski and V.~Popkov, New J. Phys. {\bf 15}, 073051 (2013).

\bibitem{groth} C. W. Groth, B. Michaelis, and C. W. J. Beenakker, Phys. Rev. B {\bf 74}, 125315 (2006); C. W. Groth, Master's Thesis, Leiden University (2007).

\bibitem{muz} B. A. Muzykantskii and Y. Adamov, Phys. Rev. B {\bf 68}, 155304 (2003); T. Antal, P. L. Krapivsky, and A. R�kos, Phys. Rev. E {\bf 78}, 061115 (2008);  M. \v Znidari\v c,  {\tt arXiv:1310.3670}.

\bibitem{TASEP} M. Gorissen \emph{et al.}, Phys. Rev. Lett. {\bf 109}, 170601 (2012); A. Parmeggiani, Physics {\bf 5}, 118 (2012) 

\bibitem{cond} D. B. Gutman, Yuval Gefen, and A. D. Mirlin, Phys. Rev. Lett. {\bf 105}, 256802 (2010).

\bibitem{cft} D. Bernard and B. Doyon, {\tt arXiv:1302.3125}.

\bibitem{mark}  Bi-Heng Liu {\em et al.}, Nat. Phys. {\bf 7}, 931 (2011).

\bibitem{qsimul} J. Simon {\em et al.}, Nature {\bf 472}, 307 (2011).

\bibitem{qsimul2} J. T. Barreiro {\em et al.}, Nature {\bf 470}, 486 (2011);  S. Hofferberth, Nat. Phys. {\bf 4}, 489-495 (2008). 

\bibitem{frust} H. T. Diep, 'Frustrated Spin Systems' (World Scientific, Singapore, 2004).

\bibitem{manzano1} D. Manzano, PloS one {\bf 8}, e57041 (2013);  D. Manzano and P. I. Hurtado, {\tt arXiv:1310.7370}.

\bibitem{majd} A. Majdandzic \emph{et al.}, Nat. Phys. {\bf10}, 34-38 (2014).

\bibitem{cap} P. Cappellaro, L. Viola and C. Ramanathan, Phys. Rev. A 83, 032304 (2011);  A. Ajoy and P. Cappellaro, Phys. Rev. Lett. {\bf 110}, 220503 (2013). 

\bibitem{third} B. Reulet, J. Senzier and D. E. Prober, Phys. Rev. Lett. {\bf 91}, 196601 (2003); Yu. Bomze {\em et al.}, {\em ibid.} {\bf 95}, 176601 (2005); G. Gershon {\em et al., ibid. } {\bf 101}, 016803 (2008).

\bibitem{scully} M.O. Scully and W.E. Lamb Jr., Phys. Rev. {\bf 179}, 368–374 (1969). 

\bibitem{levitov} L. S. Levitov and G. B. Lesovik, JETP Lett. {\bf 58}, 230 (1993); L. S. Levitov {\em et al.}, J. Math. Phys. (N.Y.) {\bf 37}, 4845 (1996); D. A. Ivanov, H. W. Lee, and L. S. Levitov, Phys. Rev. B {\bf 56}, 6839 (1997).

\bibitem{open}  H.-P. Breuer and F. Petruccione, `The Theory of Open Quantum Systems' (Oxford University Press, NY, 2002).

\bibitem{notegenform} Pure {\em dephasing} dissipators, e.g. $L_j \sim \sigma^\z_j$, commute with expr. \eqref{eq:ansatz} and do not affect results in the first order.

\bibitem{evans} D. E. Evans, Commun. Math. Phys. {\bf 54}, 293 (1977);  B. Bu\v ca and T. Prosen, New J. Phys. {\bf 14}, 073007 (2012).

\bibitem{p12} T. Prosen, Phys. Scr. {\bf 86}, 058501 (2012), and e-print: {\tt arXiv:1205.3126}.

\bibitem{noteparity} Equivalently, eigenstates of $H$ with {\em different} parities, $P\ket{\psi_{1,2}}=\pm \ket{\psi_{1,2}}$, $H\ket{\psi_{1,2}}=E_{1,2}\ket{\psi_{1,2}}$, should {\em always} have
different energies, $E_1\ne E_2$.

\bibitem{p12b} T. Prosen, Phys. Rev. Lett. {\bf 109}, 090404 (2012).

\bibitem{qt1} G.~Benenti, G. Casati, T.~Prosen, D.~Rossini and M.~\v Znidari\v c, Phys. Rev. B {\bf 80}, 035110 (2009).

%\bibitem{hub} T. Prosen,  {\tt arXiv:1310.4420};  {\tt arXiv:1310.8629}.

\end{thebibliography}
\end{document}